\documentclass{article}

\usepackage[pdftex]{hyperref}
\usepackage[utf8]{inputenc}
\usepackage{times}
\usepackage{graphicx}   
\usepackage[letterpaper, top=2 cm,bottom=2cm,
            left=2.54cm,right=2.54cm 
            ]{geometry}                
\usepackage{caption2}       
\usepackage{gensymb} 
\usepackage[version=3]{mhchem}
\usepackage[super,numbers,sort&compress]{natbib}  
\usepackage{array,multirow}
\usepackage{authblk}         
\usepackage{chemarrow}       
\usepackage{changepage}

\graphicspath{{figures/}}

\usepackage{color, soul}
\definecolor{linkColor}{rgb}{1,0,0}
\definecolor{citeColor}{rgb}{1,0,0}
\hypersetup{pdfborder={0 0 0}, colorlinks=true,urlcolor=linkColor,citecolor=citeColor}

\begin{document}
\fontsize{18pt}{24pt}\selectfont
\setlength{\parindent}{0cm}
\textbf{Strong structural and electronic coupling in metavalent PbS moir{\'e} superlattices}

\vskip 18pt
\fontsize{12pt}{14pt}\selectfont
Yu Wang$^{1,2,\dag}$, Zhigang Song$^{3,\dag}$, 
Jiawei Wan$^{1,2}$, Sophia Betzler$^{1}$,
Yujun Xie$^{1}$, Colin Ophus$^{4}$, Karen C. Bustillo$^{4}$,
Peter Ercius$^{4}$, Lin-Wang Wang$^{1}$, 
and Haimei Zheng$^{1,2,*}$

\vskip 18pt
\setlength{\parskip}{4pt}
\textbf{Affiliations:}

$^1$Materials Sciences Division, Lawrence Berkeley National Laboratory, Berkeley, CA 94720, USA.

$^2$Department of Materials Science and Engineering, University of California, Berkeley, Berkeley, CA 94720, USA.

$^3$John A. Paulson School of Engineering and Applied Sciences, Harvard University, Cambridge, MA 02138, USA

$^4$National Center for Electron Microscopy, The Molecular Foundry, Lawrence Berkeley National Laboratory, Berkeley, CA 94720, USA.

$^\dag$Contributed equally to this work.

$^*$To whom correspondence should be addressed; E-mail: hmzheng@lbl.gov

\vskip 18pt
\setlength{\parskip}{0em}
\fontsize{12pt}{18pt}\selectfont

\section*{Abstract}

Moir{\'e} superlattices are twisted bilayer materials, 
in which the tunable interlayer quantum confinement offers
access to new physics and novel device functionalities.
Previously, moir{\'e} superlattices were built exclusively using materials with weak van der Waals  interactions and synthesizing moir{\'e} superlattices with strong interlayer chemical bonding was considered to be impractical.
Here using lead sulfide (PbS) as an example,
we report a strategy for synthesizing 
moir{\'e} superlattices coupled by strong chemical bonding. 
We use water-soluble ligands as a removable template to obtain free-standing ultra-thin PbS nanosheets
and assemble them into direct-contact bilayers with various twist angles. 
Atomic-resolution imaging shows the moir{\'e} periodic structural reconstruction at superlattice interface, 
due to the strong metavalent coupling.
Electron energy loss spectroscopy and theoretical calculations collectively
reveal the twist angle-dependent electronic structure, especially the emergent separation of flat bands at small twist angles. 
The localized states of flat bands are similar to well-arranged quantum dots, promising an application in devices.
This study opens a new door to the exploration of deep energy modulations within moir{\'e} superlattices alternative to van der Waals twistronics.

\newpage
\section*{Introduction}
\setlength{\parindent}{0.5 cm}
Recently, moir{\'e} superlattices
have been synthesized by stacking two layers of
two-dimensional (2D) 
materials with relative  twist angles\cite{liu2019van,bediako2018heterointerface},
in which the long-range superlattice potentials
from interlayer interactions can create quantum confinement in each layer.
Quantum confinement in moir{\'e} superlattices can 
slow down or localize electrons, 
providing a tunable platform for studying strongly correlated physics \cite{tang2020simulation, lu2019superconductors}, 
such as superconductivity\cite{cao2018unconventional, arora2020superconductivity},
Mott insulators\cite{cao2018correlated}, 
and interacting topological insulators
\cite{nuckolls2020strongly, chen2020tunable}.
In the moir{\'e} superlattices, the 
twisting topology determines 
the 2D
quantum confinement
and it offers an additional degree of freedom to modulate the electronic structure, usually referred to as twistronics\cite{ribeiro2018twistable,hu2020topological}.
So far, all  2D moir{\'e} superlattices are synthesized  using van der Waals (vdW) materials\cite{carr2020electronic}, 
such as graphene and transition-metal dichalcogenide, 
where the two layers of materials are coupled through vdW interactions.
Twistronics based on  these  vdW  materials
has attracted great interest in various fields, ranging from physics
\cite{dean2013hofstadter, jiang2019charge, uri2020mapping, sunku2018photonic,tran2019evidence, alexeev2019resonantly, seyler2019signatures,  sharpe2019emergent}
to materials science\cite{liu2019van,sutter2019chiral,rhodes2019disorder,weston2020atomic},
and chemistry\cite{dong2018interface,kooi2020chalcogenides}. 
Different from materials coupled by chemical bonding at an interface, such as  conventional semiconductor heterostructures, vdW twistronics have reduced strength in modulating the electronic structures due to the weak interlayer coupling, although strong coupling in twistronics is desired.\cite{kazmierczak2021strain,shabani2021deep}
Most of the experimental observations of exotic electronic properties, especially those associated with electron transport, 
are realized at extremely low temperatures\cite{tang2020simulation, lu2019superconductors, cao2018correlated,nuckolls2020strongly, chen2020tunable, xu2021coexisting}. 
To increase the electronic modulation imposed by moir{\'e} superlattice, 
one approach is to replace 
the vdW interactions with strong chemical bonding such as covalent, ionic, or metavalent bonding.

Achieving strong quantum confinement in moir{\'e} superlattices 
by chemical bonding will pave a way to fabricating a new class of materials for 
beyond-vdW twistronics\cite{kennes2021moire},
and it may 
also shed light on some challenging issues in other systems. 
For example, strongly coupled moir{\'e} superlattices 
can be structurally and functionally
similar to an array of quantum dots\cite{song2020switchable}, 
offering an alternative route to super-crystals\cite{zhang2017interlayer}
by avoiding the notorious issue of connection defects formed
during quantum dot self-assembly\cite{whitham2016charge}. 
Moreover, the energy bands near the Fermi level in moir{\'e} patterns can be flattened due to the strong modulation, 
which may trap electrons in individual ``quantum-dot'' potentials
upon suitable doping, leading to Wigner crystallization
\cite{wigner1938effects, li2021imaging, regan2020mott}. 
Thus, creating  strongly coupled moir{\'e} superlattices through chemical bonding
combines the strengths  of two fields:  
the tunable confinement of 2D moir{\'e} superlattices
and the strong coupling in conventional semiconductor heterostructures.
Since  moir{\'e} superlattices cannot be achieved using  conventional semiconductor synthesis methods, 
e.g., epitaxial growth\cite{liu2019van,oh2020design}, 
it is unclear  whether it is possible to synthesize  chemically bonded moir{\'e} superlattices.

Here, we use PbS as a model system to demonstrate a strategy for 
constructing moir{\'e} superlattices with strong interlayer coupling through metavalent bonding. 
Such beyond-vdW moir{\'e} superlattices are obtained for the first time through chemical synthesis.
Strong interfacial coupling of the superlattice is revealed through 
atomic-resolution imaging, and the giant electronic modulation of  moir{\'e} pattern
is validated through the combination of electron energy loss spectroscopic analysis and theoretical calculations.

\section*{Results and Discussion}\label{results}

\subsection*{Conceptual discussion of metavalent moir{\'e} superlattices}

\begin{figure}[htb]
	\begin{center}
		\includegraphics[width=1\textwidth]{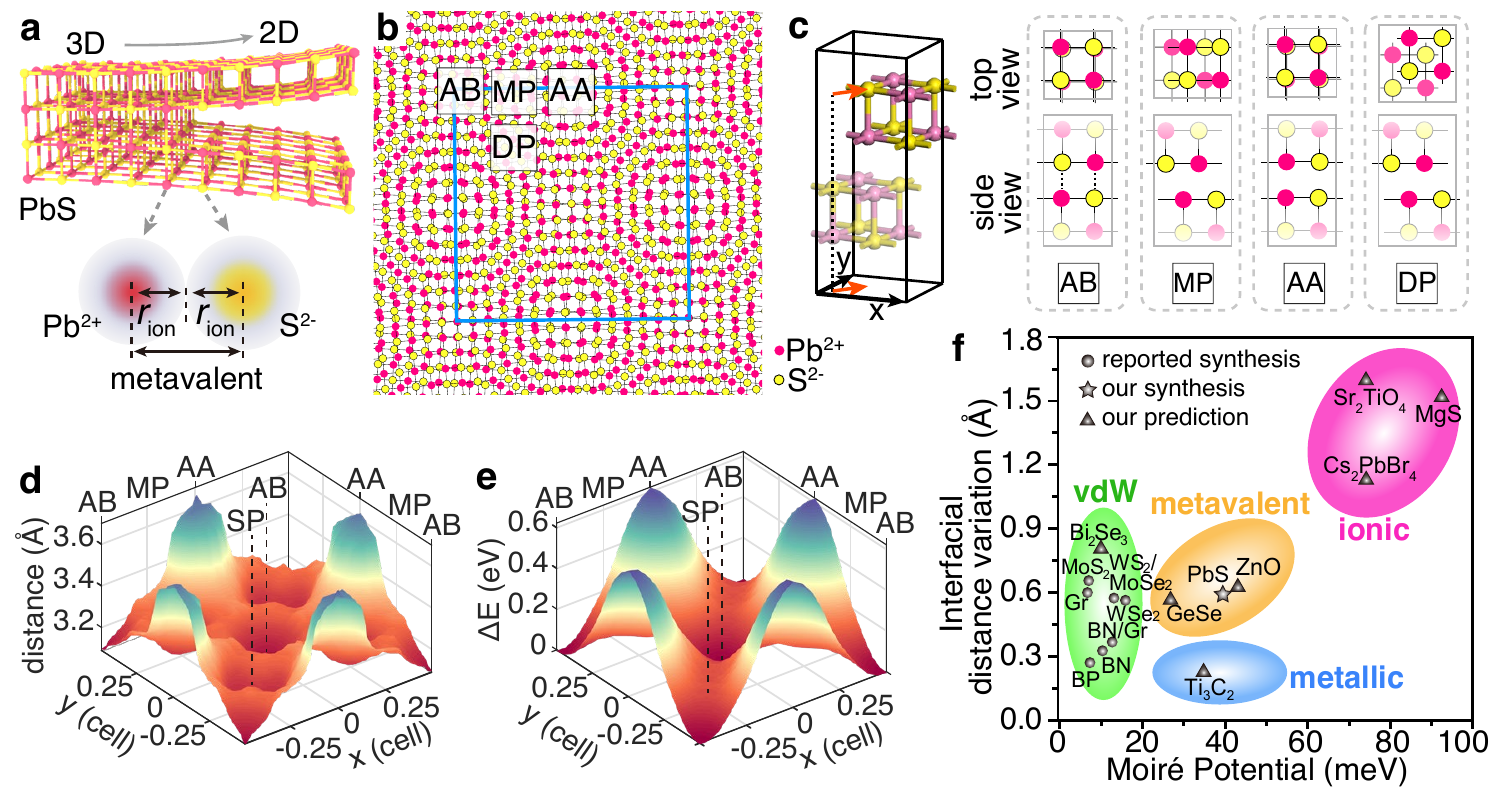}
	\end{center}
	\caption{\textbf{Structure and strong coupling of PbS moir{\'e} superlattice}.
    \textbf{a}, 3D and the cleaved 2D structures of PbS rock-salt crystal, emphasizing the metavalent interaction in all directions.
    \textbf{b}, Different local atomic alignments occur in a PbS moir{\'e} superlattice with a twist angle of 8\degree. Blue square marks the moir{\'e} unit-cell. Four representative stacking configurations are  highlighted as Pb on Pb (AA), Pb on S (AB), middle point (MP), and diagonal point (DP).
    \textbf{c}, Illustration of laterally shifting two PbS nanosheets for creating different stacking configurations. Top views and side views of representative structures are exemplified.
    \textbf{d},\textbf{e}, DFT calculations of the structures created according to panel (c) showing the interfacial distance ({d}) and the free energy change ({e}) upon the bilayer lateral shift.
    \textbf{f}, DFT calculations on  varieties of moir{\'e} superlattices, including the reported vdW superlattices, the metavalent PbS synthesized in this work, and our predictions of other chemically bonded superlattices.
    The results show their moir{\'e} potential and the largest interfacial distance variation among different stacking configurations.
    }
	\label{fig:main:concept}
\end{figure}

We first describe  theoretically why metavalent moir{\'e} superlattices 
can give rise to stronger coupling effects than  vdW moir{\'e} superlattices. 
As shown in Fig. \ref{fig:main:concept}a, bulk PbS has a  rock-salt crystal structure 
and features an unconventional chemical bonding between Pb and S atoms, namely metavalent bonding,
in which the valence electrons are delocalized to an extent between covalent and metallic bonding\cite{kooi2020chalcogenides}. 
This metavalent Pb--S bonding is much stronger than vdW interactions\cite{liu2019van},
and it can be used as the interlayer interaction to construct strongly coupled PbS moir{\'e} superlattices 
if one can assemble ultra-thin 2D PbS nanosheets with pristine surfaces into twisted bilayers.
This methodology may be further generalized to obtain new moir{\'e} superlattices 
coupled by other types of interfacial chemical bonding, for example, 
{MXene}\cite{naguib201425th}
moir{\'e} superlattices coupled by metallic bonding, 
perovskite\cite{ji2019freestanding}
moir{\'e} superlattices coupled by ionic bonding,
and twisted bilayers of metal--organic frameworks\cite{wu2016conetronics}
coupled by mixed interactions.

PbS moir{\'e} supercells with small commensurate angles have a tetragonal
symmetry (Fig. \ref{fig:main:concept}b), 
in which two types of interfacial atoms stacked into four general configurations: 
Pb on Pb 
(marked as AA), Pb on S 
(marked as AB), 
the middle point (MP) between AA and AB,
and the diagonal point (DP) between two AB positions.
In small-angle twisted bilayers, each stacking configuration can be approximated by small unit cells consisting of laterally shifted bilayers,
in which the shift coordinates are scaled by $x$ and $y$ in the unit of cell size 
(Fig. \ref{fig:main:concept}c).
Therefore, the AB, AA, MP, and DP configurations correspond to the ($x$, $y$)  of
(0, 0), (0, 0.5), (0, 0.25), and (0.25, 0.25), respectively, 
and other configurations between the four extrema
can be constructed with continuous ($x$, $y$) shift.

Density-functional theory (DFT) calculations on the created structures (detailed in Supplementary Text S1) suggest that different stacking configurations can lead to deep modulations of interfacial reconstruction and electronic properties.
Fig. \ref{fig:main:concept}d,e shows the changes of  interlayer distance and free energy  as functions of lateral interlayer shift. 
The surface plot regarding  either interlayer distance or free energy
has the maxima at AA spots and the minima at AB spots. 
The difference between the maximum and minimum interlayer distance is as large as 0.6 \AA, implying a possible structural reconstruction at the interface of a small-angle twisted  superlattice. 

Maximal free energy fluctuation in the real space is
defined as moir{\'e} potential\cite{tran2019evidence, halbertal2021moire},
which is an important indicator of the strength of energy modulation 
for a given moir{\'e} superlattice.
We estimate the moir{\'e} potential of PbS and other 2D materials by calculating
the largest free energy difference among all possible stacking configurations
using approximate small unit cells. 
Fig. \ref{fig:main:concept}f shows the calculated moir{\'e} potentials of various materials, 
including the reported vdW superlattices, the metavalent PbS synthesized in this work,
and our predictions of other chemically bonded superlattices.
The moir{\'e} potential of PbS is 
40 meV per atom, more than twice of the reported vdW superlattices.
Generally,
chemical  bonding (e.g., metallic, metavalent, and ionic bonding)  leads  to much deeper energy modulation compared to vdW interactions. 
The deep energy modulation can localize electrons in the high-symmetry points with local energy extrema, 
providing an array of identical quantum-dot-like potentials\cite{zhang2017interlayer,song2020switchable}.
Until now, the properties of the chemically bonded moir{\'e} superlattices 
and their structural stability remain  unknown
due to the lack of a synthesis strategy.


\subsection*{Synthesis and characterization of PbS moir{\'e} superlattices}

\begin{figure}[p]
	\begin{center}
		\includegraphics[width=0.8\textwidth]{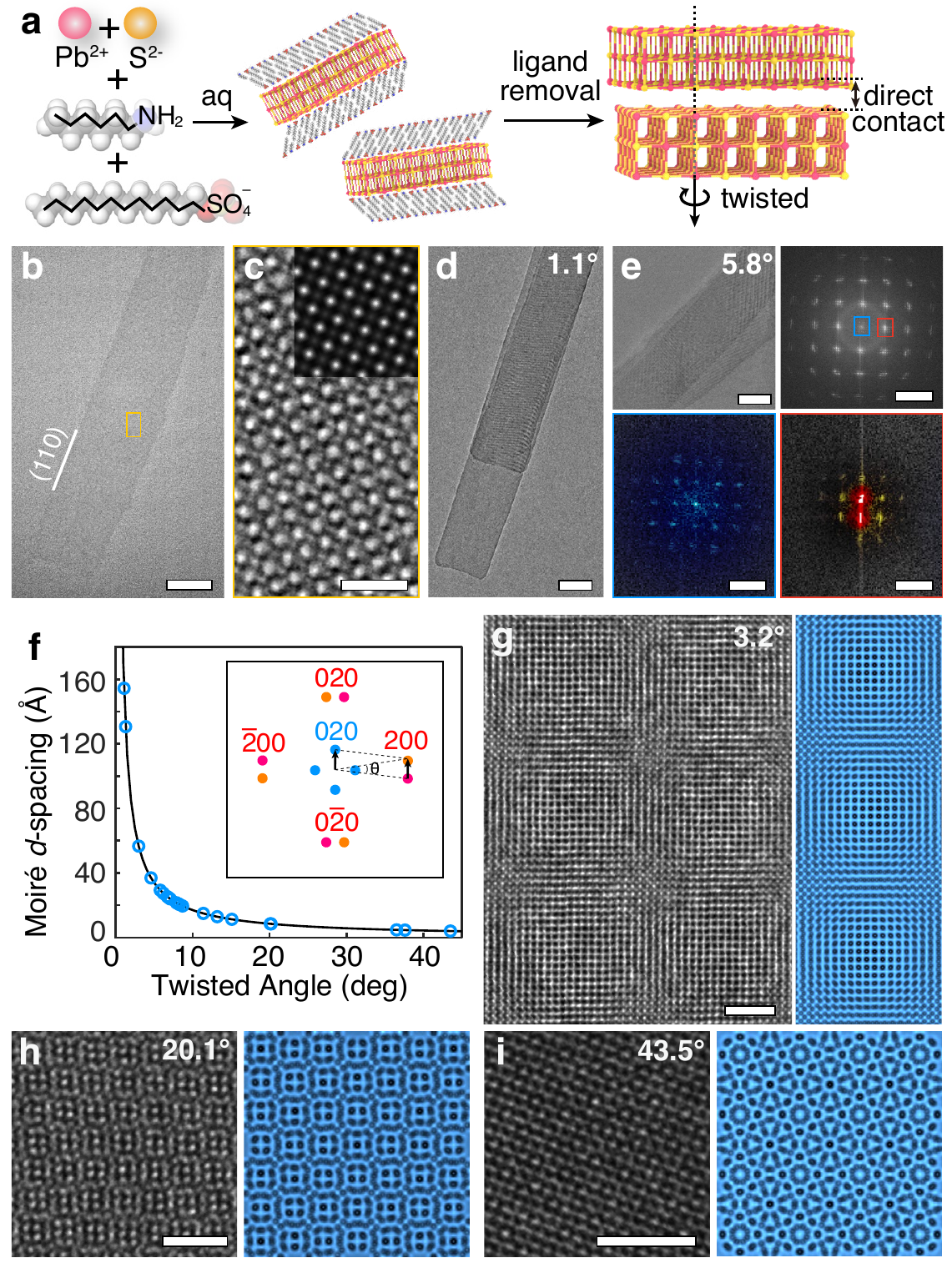}
	\end{center}
	\caption{\textbf{Synthesis and TEM characterizations of PbS moir{\'e} superlattices}.
    \textbf{a}, Schematics of synthetic procedures. 
    \textbf{b}, Low magnification  TEM image of a single ultra-thin PbS nanosheet. 
    \textbf{c}, Atomic-resolution image of the boxed area in panel b with an inserted simulation of TEM image. 
    \textbf{d}, TEM image of a bilayer moir{\'e} superlattice   with a 1.1\degree~twist angle.
    \textbf{e}, TEM image (top-left) and FFT pattern (top-right) of a bilayer moir{\'e} superlattice with a 5.8\degree~twist angle. Two bottom panels show the enlarged details of the FFT pattern, in which blue dots indicate moir{\'e} spatial frequencies, red dots indicate the (200) spatial frequencies of two individual rock-salt nanosheets, and orange dots indicate the emerged pattern from moir{\'e} pattern and individual rock-salt patterns.
    \textbf{f}, Theoretical (black curve) and observed (green hollow dots) relationship between moir{\'e} $d$-spacing (200) and twist angles. Inset shows the relationship between moir{\'e} spatial frequencies (green) and two sets of individual rock-salt spatial frequencies (red and orange).
    \textbf{g}--\textbf{i}, Atomic-resolution TEM images and corresponding simulated images (false coloured) of  bilayer moir{\'e} superlattices with a variety of twist angles.
    Scale bar: b, 30 nm; c, 1 nm; d, 100 nm; e, 30 nm (top left), 5 nm$^{-1}$ (top right), 0.5 nm$^{-1}$ (bottom two); g--i, 2 nm.
    }
	\label{fig:main:synth-TEM}
\end{figure}

We use metavalent PbS as a model system of beyond-vdW  moir{\'e} superlattices to assess the feasibility of achieving
the predicted modulation of electronic structures.
Ultra-thin PbS nanosheets have been previously synthesized in organic low-polar solvents\cite{schliehe2010ultrathin},
in which the interaction between the solvent-phobic PbS core and the long-alkyl-chain ligands is designed to be strong to  guide asymmetric growth and stabilize the formed nanocrystals.
However, the strong core--ligand interaction also leads to difficulty in ligand removal,
and the ligands prevent direct metavalent bonding
between two nanosheets\cite{schliehe2010ultrathin}. 
To overcome this dilemma, we developed an aqueous synthesis strategy employing two surfactant ligands that have adequate solubility and bind moderately 
with the inorganic core.
The schematic in Fig. \ref{fig:main:synth-TEM}a shows that
\ce{Pb^2+} and \ce{S^2-} precursors and two organic ligands (i.e., hexylamine and dodecyl sulfate) in 
an acidic aqueous solution at 80\celsius~for 20 min produce ligand-capped ultra-thin PbS nanosheets (Supplementary Method S1). 
TEM imaging and statistics (Supplementary Fig. S1) show that the nanosheets have a rectangular shape with an average width and length of around 40 nm × 200 nm. 
The synthetic mechanism is discussed in Supplementary Text S2.
Due to the high polarity of the PbS surfaces, high-polar solvents (such as water) are required
to remove the ligands. 
In this synthesis method, both ligands can be readily removed by washing with dilute basic and acidic aqueous solutions alternatively,
in contrast to the long-alkyl-chain ligands used in conventional synthesis. 
After ligand removal, the naked PbS nanosheets are immediately drop-casted, allowing the assembly of moir{\'e} superlattices
through solvent evaporation.

Aberration-corrected transmission electron microscopy (TEM) imaging (Fig. \ref{fig:main:synth-TEM}b)  and the 
zoomed-in portion from the yellow box (Fig. \ref{fig:main:synth-TEM}c) with
corresponding image simulation 
(Fig. \ref{fig:main:synth-TEM}c inset) 
show that the as-synthesized nanosheets have a rock-salt structure with $\{$001$\}$ surfaces and $\{$110$\}$ edges. 
Energy-dispersive X-ray spectroscopy  confirms that the moir{\'e} superlattice consists of Pb and S atoms at a molar ratio of 1:1 (Supplementary Fig. S2).
Bilayer moir{\'e} superlattices with various twist angles are observed at low magnification due to the presence of moir{\'e} fringes (Fig. \ref{fig:main:synth-TEM}d,e).
A fast Fourier transform (FFT) of a representative moir{\'e} superlattice TEM image shows the expected pattern of two rotated sets of spatial frequencies corresponding to the structure of  each sheet (Fig. \ref{fig:main:synth-TEM}e). 
In the enlarged images (bottom panels of Fig. \ref{fig:main:synth-TEM}e), the two red dots correspond to the
(200) $d$-spacing of two individual sheets, the set of blue dots  correspond to the  $d$-spacings of the moir{\'e} pattern, 
and the orange dots correspond to the addition of the spatial frequencies
from individual sheets and the moir{\'e} pattern.
The theoretical relationship between the FFT pattern of an individual nanosheet and that of a moir{\'e} pattern
is illustrated in  Fig. \ref{fig:main:synth-TEM}f inset, 
and accordingly, 
the moir{\'e} $d$-spacing can be calculated by: 
    \begin{equation}
    \label{eq-dspacing}
        d{\rm _{m}}(200) = 
         \frac{d{\rm _{rs}}(200)}{2\cdot \sin(\theta / 2)}
    \end{equation}
where $\theta$ is the twist angle; $d{\rm _{m}}$ and $d{\rm _{rs}}$ are the (200) $d$-spacing of the moir{\'e} cell and the rock-salt cell, respectively.
In addition, the moir{\'e} $d$-spacing can also be directly measured from high-resolution TEM images for a variety of twist angles, 
verifying the calculated results from Eq. \ref{eq-dspacing} as plotted in Fig. \ref{fig:main:synth-TEM}f.

Fig. \ref{fig:main:synth-TEM}g--i and Supplementary Fig. S3
show  atomic-resolution TEM images and  corresponding image simulations (blue images; see Supplementary Method S2) of moir{\'e} superlattices with various twist angles.
For smaller twist angles,
the (quasi-) unit cell of the moir{\'e} pattern is more obvious, and the superlattice appears as an array of identical quantum dots.
For example, the moir{\'e} superlattice with a 3.2\degree~
twist angle (measured by the rotation angle of FFT spots) 
consists of periodic AA/AB regions arranged in square symmetry 
with DP regions filled in the diagonal positions (Fig. \ref{fig:main:synth-TEM}g),
resembling an epitaxially fused superlattice of 5$\sim$6 nm PbS quantum dots\cite{whitham2016charge}.
At a large twist angle close to 45\degree,
the moir{\'e} superlattice resembles 2D octagonal quasicrystals  (Fig. \ref{fig:main:synth-TEM}i),
showing 
an approximate {\emph{C}$_8$} symmetry in TEM image and reflecting the {\emph{S}$_8$} symmetry of octagonal quasicrystals. 

\subsection*{Structural analysis of moir{\'e} superlattice interface}

\begin{figure}[p]
	\begin{center}
		\includegraphics[width=1\textwidth]{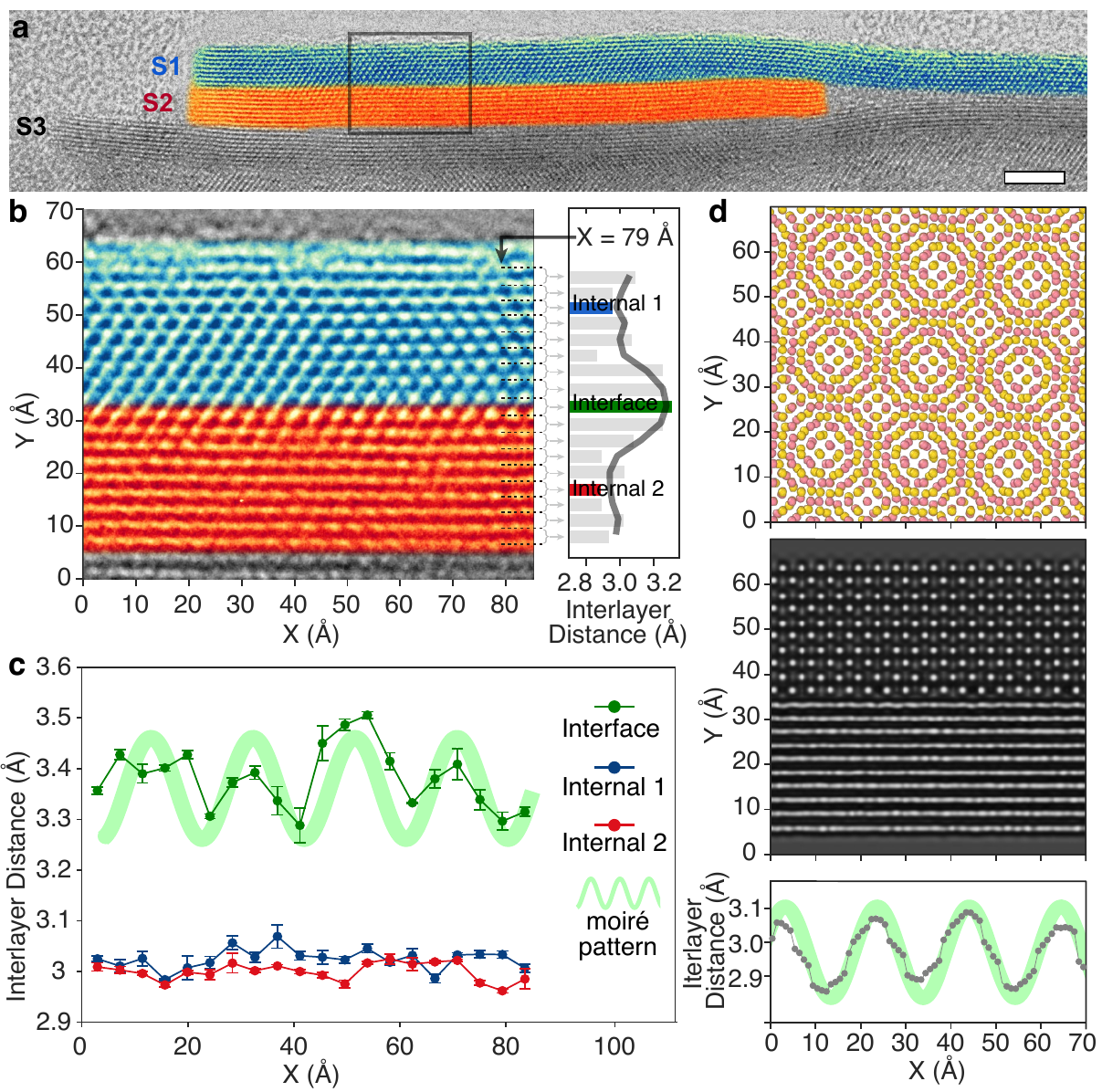}
	\end{center}
	\caption{\textbf{Structural analysis of moir{\'e} superlattice interface}.
    \textbf{a}, Side view of a moir{\'e} superlattice composed of three sheets, S1, S2, and S3.
    S1 and S2 are false-coloured for clarity. Scale bar, 5 nm.
    \textbf{b}, Higher-resolution image of the boxed region in panel a and  
    the demonstration of measuring interlayer distances.
    The Y positions of each layer (marked by dash lines) are evaluated by finding the brightest dots/strips in the image at the section of X = 79 \AA,
    and the interlayer distances are calculated as the difference of neighbouring Y positions
    (marked by brackets and arrows). 
    Two internal interlayers in each sheet and the interface of S1 and S2
    are highlighted by blue, red, and green, respectively, as representative interlayers
    for further analysis in panel c.
    \textbf{c}, Interlayer distance fluctuation by measuring three representative interlayers
    at different X locations. Error bars indicate the standard deviation of  measurements using three different sampling widths of 4, 6, and 8 \AA. 
    Thick green line shows the moir{\'e} periodicity of the bilayer superlattice 
    with an 11.5\degree~twist angle.
    \textbf{d}, Theoretically optimized atomic structure (up), simulated side-view image (middle), and 
    calculated interfacial distance fluctuation (bottom, gray dots)
    of an 11.5\degree~twist superlattice. 
    Moir{\'e} periodicity (thick green line) for comparison.
    }
	\label{fig:main:sideview-TEM}
\end{figure}

Fig. \ref{fig:main:sideview-TEM}a shows the TEM image along the basal plane (side-view) of a moir{\'e} superlattice composed of three PbS sheets (labelled as S1, S2, and S3).
The three sheets have a consistent thickness, approximately 3.0 nm,
and consist of 10 (002) planes.
Neighbouring sheets (S1--S2 or S2--S3) are in direct contact with 
3 \AA~interfacial spacing and
no trace of ligands at interface.  
For comparison, a bilayer with even one interfacial ligand residue would appear rather different, 
showing a larger interfacial distance over 5 \AA~(Supplementary Fig. S4).
The clear interface between PbS sheets  suggests 
the efficiency of the as-described ligand removal method. 

Note that, S1 and S3 are deformed beyond the right end of S2, indicating strong interlayer metavalent interactions and large deformability of naked PbS at the sub-10-nm scale\cite{wang2019dynamic}.
This large deformability of uncoupled nanosheets
is consistent with the self-rolling behaviour 
of individual naked nanosheets in the solution phase (Supplementary Text S3).
Additionally, this also  suggests that forming superlattices is an effective way to stabilize the naked nanosheets.

Further scrutiny of the side-view image in Fig. \ref{fig:main:sideview-TEM}b reveals that S1 is oriented along a [110] axis regarding the viewing direction based on the observed elongated hexagonal pattern, whereas S2 is tilted away from a low-index zone axis. 
This indicates that the sheets are rotated around the basal plane regarding one another, confirming the formation of moir{\'e} superlattices with direct interlayer contact. 
We further measured the lattice spacings in the superlattice at different X locations in real space based on TEM image simulations and the recognition of the image peak positions, 
as detailed in Supplementary Text S4. 
The right graph in Fig. \ref{fig:main:sideview-TEM}b shows
an example at X = 79 \AA,
where the (002) spacings are approximate to 3.0 \AA~(similar to that in the bulk crystal) for the inner layers inside each sheet,
but become slightly larger at the S1--S2 interface.
We select two internal interlayers in each sheet (labelled as Internal 1 and 2)
and the S1--S2 interface representative interlayers,
and examine their interlayer spacing at different X locations.
As shown in Fig. \ref{fig:main:sideview-TEM}c,
the spacings of Internal 1 and 2 are consistent to \textit{c.a.} 3.0 \AA~
with small fluctuation less than 0.1 \AA,
however, both the mean value (3.35 \AA)  and the fluctuation (over 0.2 \AA)
of interfacial spacing are
considerably larger.
Moreover, the fluctuation of interfacial spacing exhibits
a periodicity that matches with the theoretical 
moir{\'e} periodicity of bilayer superlattice 
with a 11.5\degree~twist angle.
We simulate the side-view image of the theoretically optimized
11.5\degree~twist superlattice and calculate
the interfacial spacing in the simulated image (Fig. \ref{fig:main:sideview-TEM}d).
There is indeed a considerable interfacial spacing fluctuation
in the simulated image that  matches with the theoretical 
moir{\'e} periodicity.
The experimental and theoretical structural analysis
jointly suggests the structural reconstruction at superlattice interface
and its correlation to moir{\'e} periodicity.

\subsection*{Twist angle-dependent electronic states of moir{\'e}  superlattices}

\begin{figure}[b!]
	\begin{center}
		\includegraphics[width=0.8\textwidth]{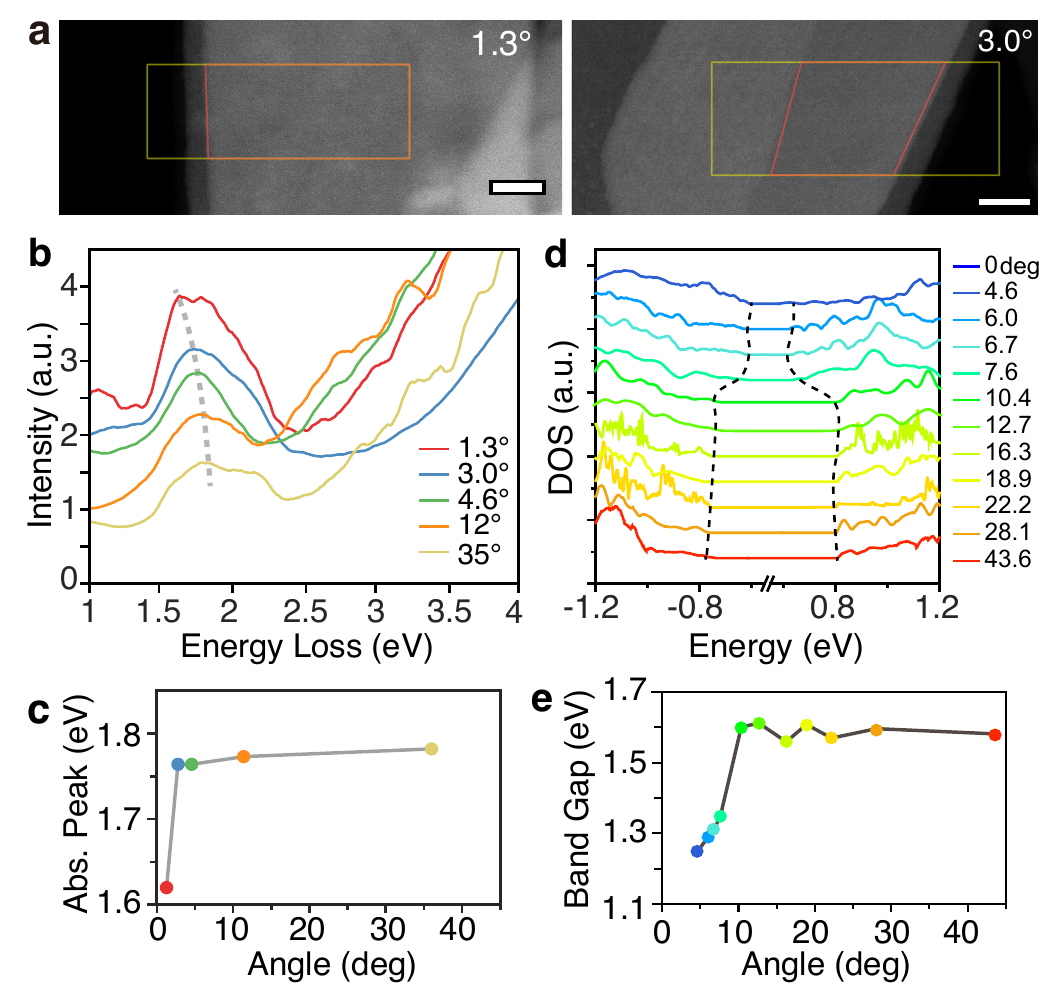}
	\end{center}
	\caption{\textbf{Twist angle-dependent STEM-EELS and electronic states of moir{\'e} superlattices}.
    \textbf{a}, Overview images of two representative superlattices with twist angles of 1.3\degree~and 3.0\degree. 
    Yellow and red boxes mark the scanning regions and bilayer regions for STEM-EELS integration, respectively.
    Scale bar, 20 nm.
    \textbf{b}, Energy loss spectra of the bilayer regions of  moir{\'e} superlattices with a variety of twist angles. Energy dispersion of all spectra is 9 meV/pixel.
    \textbf{c}, Change of spectral peak  upon twist angles extracted from panel d.
    \textbf{d}, Direct calculation of the density-of-states of commensurate moir{\'e} superlattices with a variety of twist angles. Black dashed lines connect the positions of first valence or conduction band for each structure.
    \textbf{e}, Bandgap change upon twist angles extracted from panel d.
    }
	\label{fig:main:EELS}
\end{figure}

We employ monochromated electron energy loss spectroscopy (EELS) in an aberration-corrected scanning TEM (STEM) to investigate the
electronic excitation of moir{\'e} superstructures as a function of twist angle.
Owing to  the advantage of the latest direct detection camera and the large absorption efficiency of PbS, the low-loss spectra in the exciton region exhibit high signal-to-noise ratios 
(Supplementary Fig. S5). 
Fig. \ref{fig:main:EELS}a shows the overview images
of two representative superlattices with  twist angles of 1.3\degree~and 3.0\degree~
and the corresponding EELS scanning profiles.
Other superlattices that were measured have larger twist angles 
4.6\degree, 12\degree, and 35\degree.
Because all  moir{\'e} superlattices  were measured in one experiment with identical conditions, we can compare the thickness of individual sheets from 
the scattering intensity from non-overlapping regions with only a single sheet in projection.
We find that all individual nanosheets in the moir{\'e} superlattices have a similar thickness with less than 10\% difference.

The integrated spectra of the double-layer regions with
various twist angles are plotted
in Fig. \ref{fig:main:EELS}b, and the peak locations are extracted
in Fig. \ref{fig:main:EELS}c.
These plots show that the spectral peak slightly shifts to lower energy as the twist angle decreases from 35.3\degree~to 3.0\degree, but the peak energy dramatically decreases from  3.0\degree~to 1.3\degree.

Features in low-loss EEL spectra  arise due to inter-band excitation and intra-band transitions in a similar way to optical spectra\cite{hage2018nanoscale,gogoi2019layer},
which are qualitatively reflective of 
the electronic excitation or the bandgap.
To understand the abrupt EELS peak change at small twist angle,
we perform direct self-consistent calculations by constructing large moir{\'e} unit cells (up to 4360 atoms) with commensurate angles (Supplementary Text S1).
The density of states (DOS) of the moir{\'e} cells with twist angles ranging from 4.47\degree~to 43.6\degree~are shown in Fig. \ref{fig:main:EELS}d, 
and the corresponding bandgaps are plotted in Fig. \ref{fig:main:EELS}e.
For superlattices with a twist angle larger than 10.4\degree, the bandgap fluctuates around a constant. However, the bandgap drops rapidly as the twist angle decreases 
from 10.4\degree~to smaller angles (Fig. \ref{fig:main:EELS}e).
This calculated relationship between bandgap and twist angle,
especially the sharp drop of bandgap at a small angle,
is in good agreement with our EELS  experiments (Fig. \ref{fig:main:EELS}c).
The experimental and theoretical results collectively 
suggest that the band structure of metavalent moir{\'e} superlattices, is highly dependent on the twist angle, especially at small twist angles.

\subsection*{Band structure calculations of PbS moir{\'e} superlattices}

\begin{figure}[ht]
	\begin{center}
		\includegraphics[width=0.75\textwidth]{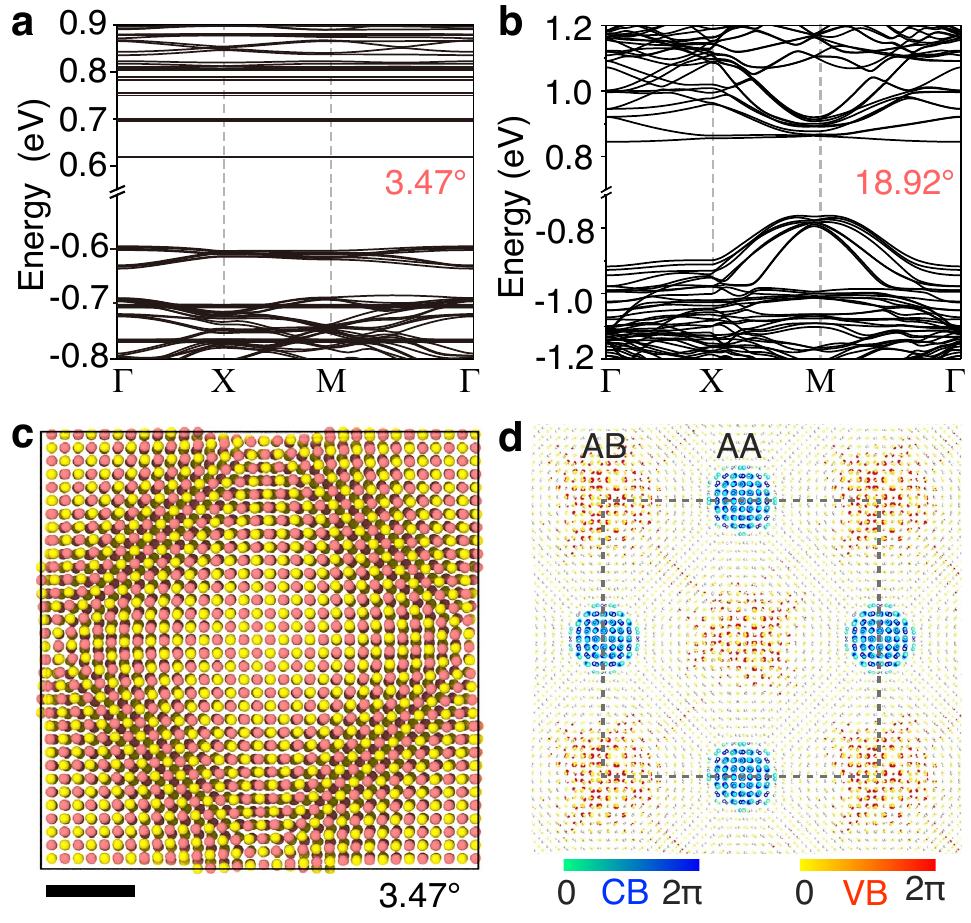}
	\end{center}
	\caption{\textbf{DFT calculations of the electronic localization}.
    \textbf{a},\textbf{b} Calculated band structures of the superlattices with twist angles of 3.47\degree~and 18.92\degree.
    \textbf{c}, Theoretically optimized moir{\'e} cell of the superlattice with a twist angle of 3.47\degree. 
    \textbf{d}, Charge density of valence band (VB) and conduction band (CB). Colour is coded by the phase of wave function. Moir{\'e} cell is marked by the dashed line.  
    }
    \label{fig:main:DFT}
\end{figure}

In addition to the twist angle-dependent bandgap,
the DFT calculations of band structures 
reveal the emergent separation of electronic states of superlattices
at small twist angles (Fig. \ref{fig:main:DFT}a,b and
Supplementary Fig. S6).
As the twist angle of a  superlattice falls below 10\degree,
the bands become narrower due to the quantum containment of the moir{\'e} pattern, and
bands near the Fermi level begin to separate from the deep and high-energy states.
For example, in the case of 3.47\degree,
the bands near the Fermi level, especially the conduction band, become extremely flat (Fig. \ref{fig:main:DFT}a,b). 
This emergent separation of electronic states is probably the reason of
the observed peak split in the EELS spectrum at 1.3\degree~(Fig. \ref{fig:main:EELS}b).
Supplementary calculations 
(Supplementary Fig. S6)
show that the separation of moir{\'e} bands,
which reflects  the modulation effect by moir{\'e} pattern,
becomes weaker as nanosheets  become thicker.
This separation of moir{\'e} bands, 
which reflects  the modulation effect by moir{\'e} pattern,
becomes weaker as nanosheets  become thicker. 
This thickness-dependent of  modulation 
indicates the value of synthesizing ultra-thin free-standing nanosheets,
stimulating future  developments in the chemical and/or physical syntheses 
of ultra-thin beyond-vdW nanosheets 
and their strongly coupled moir{\'e} superlattices.

Furthermore, these separated moir{\'e} bands at small twist angles are highly spatially localized. 
Fig. \ref{fig:main:DFT}d shows 
the corresponding wave functions of the conduction and valence bands, which are localized at  AB and AA spots, respectively, with diameters of a few nanometres.
This also indicates that
the PbS moir{\'e} superlattice not only structurally resembles but also functionally mimics
a well-arranged array of quantum dots
with separate energy levels and electron orbitals. 
Our additional calculation (Supplementary Text S5) suggests that the strong metavalent modulation by the moir{\'e} pattern may lead to  emergent opto-electronic  properties in valleytronics, 
such as valley-dependant optical selection rules.

\section*{Conclusion}
We have established an approach to introduce strong metavalent interlayer interactions into moir{\'e} superlattices. 
Our  synthesis strategy and findings on the chemically bonded moir{\'e} superlattices extend the current twistronics. 
Combining structural analysis, EELS measurements, and DFT calculations,
we demonstrate that strong moir{\'e} modulation at a small twist angle
can lead to considerable structural reconstruction and electronic renormalization.
This study  provides a route to arrays of identical ``quantum-dot'' potentials by achieving deep energy modulation through metavalent interactions, providing an alternative platform for spatially variant electronic and opto-electronic  properties. We anticipate that further experimental and theoretical studies on beyond-vdW moir{\'e} superlattices will find 
more types of interlayer interactions that may result in 
strong electronic coupling, strong correlation, and realizing tunable emergent quantum properties.

\section*{Data availability}
The data that support the findings of this study are available from the
corresponding authors on reasonable request.

\bibliography{bibliography}
\bibliographystyle{naturemag}

\subsection*{Acknowledgements}
The work was supported by the U.S. Department of Energy (DOE), Office of Science, Office of Basic Energy Sciences (BES), Materials Sciences and Engineering Division under Contract No. DE-AC02-05-CH11231 within the KC22ZH program.
Y.W. was partially supported by the UC Office of the President under the UC Laboratory Fees Research Program Collaborative Research and Training Award LFR-17-477148.
Work at the Molecular Foundry was supported by the Office of Science, Office of Basic Energy Sciences, of the U.S. Department of Energy under Contract No. DE-AC02-05CH11231.

\subsection*{Author contributions}
Y.W. conceived and H.Z. supervised this project.
Y.W. and J.W. designed and performed the synthesis.
Y.W. performed TEM imaging with contribution from Y.X.
P.E. and Y.W. conducted EELS measurement with 
contribution from S.B. and K.B.
C.O. provided the code for analysing side-view images.
Y.W. analysed all experimental data.
Z.S. performed theoretical analysis 
under the supervision of L.-W.W.
Y.W., Z.S., and H.Z. wrote the manuscript with input from all authors.

\subsection*{Competing interests}
The authors declare no competing interests.

\subsection*{Supplementary information}
Supplementary Figs. S1--S5, Supplementary Methods, and Supplementary Text S1--S5 are available for this paper at https://doi.org/.

\end{document}